\documentclass[twocolumn,prb,aps,10pt]{revtex4-1}

\usepackage{amsfonts,amssymb}
\usepackage[sumlimits,intlimits]{amsmath}
\usepackage{graphics}
\usepackage{graphicx}
\usepackage{epsfig}
\usepackage{mathrsfs}
\usepackage{textcomp}
\usepackage{verbatim}
\usepackage{hyperref}    
\usepackage{bm}          

\newcommand{\be}{\begin{equation}}
\newcommand{\ee}{\end{equation}}
\newcommand{\sgn}{\operatorname{sgn}}
\newcommand{\e}{\varepsilon}
\newcommand{\rr}{\vec{r}}
\newcommand{\eq}[1]{Eq.\ (\ref{eq:#1})}
\newcommand{\avg}[1]{\langle #1 \rangle}
\newcommand{\pt}{\widetilde{\phi}}
\newcommand{\CA}{Cd$_3$As$_2$ }

\begin{document}

\title{Coulomb disorder in three-dimensional Dirac systems}

\author{Brian Skinner}
\affiliation{Materials Science Division, Argonne National Laboratory, Argonne, IL 60439, USA}

\date{June 9, 2014}

\begin{abstract}

In three-dimensional materials with a Dirac spectrum, weak short-ranged disorder is essentially irrelevant near the Dirac point.  This is manifestly not the case for Coulomb disorder, where the long-ranged nature of the potential produced by charged impurities implies large fluctuations of the disorder potential even when impurities are sparse, and these fluctuations are screened by the formation of electron/hole puddles.  In this paper I present a theory of such nonlinear screening of Coulomb disorder in three-dimensional Dirac systems, and I derive the typical magnitude of the disorder potential, the corresponding density of states, and the size and density of electron/hole puddles.  The resulting conductivity is also discussed.
\end{abstract}
\maketitle

\section{Introduction}
\label{sec:intro}

Due to its long-ranged nature, Coulomb disorder often has dramatic consequences even in situations where short-ranged disorder does not. Thus, for example, Coulomb disorder has always deserved special consideration in the physics of semiconductors,\cite{shklovskii_electronic_1984} and such studies have revealed a great number of diverse and interesting scientific phenomena over the preceding half-century.

The large qualitative difference between short-ranged and Coulomb disorder is particularly pronounced for three-dimensional (3D) Dirac materials, in which the electron kinetic energy $\e$ is linearly proportional to the momentum $\hbar \vec{k}$ according to $\e = \hbar v |\vec{k}|$.  To see this difference between short-ranged and Coulomb disorder qualitatively, one can compare the behavior of long-wavelength (small-$k$) electron states in a 3D Dirac system (3DDS) for the two cases.  Suppose, for example, that the 3DDS has some concentration $N$ of positive and negative impurities per unit volume, each with random position and random sign, and that these are taken to be either short-ranged, with finite range $a$ and typical potential $\pm V_0$, or Coulomb, with charge $\pm e$.  In the short-ranged case, an electron wavepacket with size $\lambda \gg N^{-1/3} \gg a$ experiences disorder from $\sim N \lambda^3$ impurities.  The average value of the disorder potential created by these impurities is zero, since impurities with opposite signs are equally plentiful, but statistical fluctuations in the impurity concentration create a typical excess of $\sim \sqrt{N \lambda^3}$ impurities with one of the two signs.  Thus, the volume-averaged disorder potential experienced by the electron is $\sim V_0 \sqrt{N \lambda^3}/(\lambda/a)^3 \propto 1/\lambda^{3/2}$.  The electron kinetic energy, on the other hand, scales as $\e \propto k \propto 1/\lambda$.  One can therefore conclude that short-ranged disorder has a perturbatively small effect on the electron energy for large-wavelength electron states (i.e., for states close to the Dirac point).  This robustness of the 3D Dirac point against short-ranged disorder has long been understood theoretically,\cite{goswami_quantum_2011, hosur_charge_2012, kobayashi_density_2014, syzranov_critical_2014, sbierski_quantum_2014, ominato_quantum_2014} and consequently ``Dirac semimetal" phases with vanishing density of states (DOS) are generally predicted to survive short range disorder.
(In fact, a very recent paper \cite{nandkishore_rare_2014} has shown that rare resonances between short-ranged impurities can create an exponentially small DOS at the Dirac point.)

\begin{figure}[htb!]
\centering
\includegraphics[width=0.45 \textwidth]{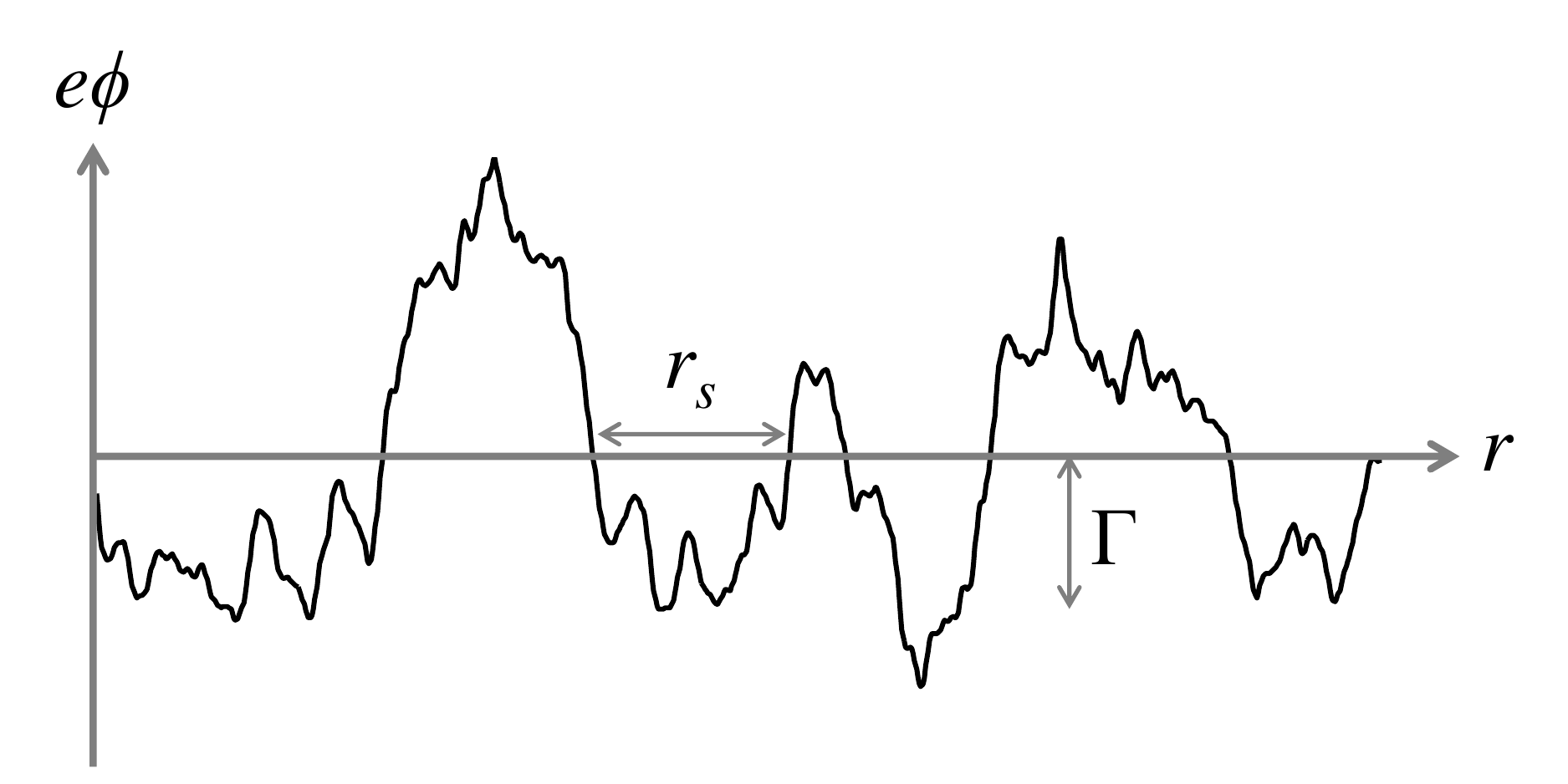}
\caption{Schematic illustration of electron/hole puddles in a 3DDS at the Dirac point.  The Coulomb potential energy $e \phi$ is shown as a function of some coordinate $r$.  Coulomb impurities create potential fluctuations with typical size $r_s$ and amplitude $\Gamma$.  Regions of positive $e \phi$ correspond to electron puddles, while negative $e \phi$ implies hole puddles.}
\label{fig:disorder-schematic}
\end{figure}

Now consider the case of disorder produced by long-ranged Coulomb impurities.  As before, an electron wavepacket with large size $\lambda$ encloses many impurities of both signs, and the net charge of these is $\sim \pm e \sqrt{N \lambda^3}$.  If one naively calculates the potential energy created by these impurities, one finds that the typical Coulomb potential energy experienced by the electron is $\sim e^2 \sqrt{N \lambda^3}/\kappa \lambda$ (in Gaussian units), where $\kappa$ is the dielectric constant.  Thus, the disorder potential \emph{grows} with increasing wavelength as $\lambda^{1/2}$, rather than falling off quickly and becoming irrelevant.  Clearly, such Coulomb impurities must have a large and nonperturbative effect near the Dirac point at any finite concentration.  As one might expect, the growth of the Coulomb potential at large length scales is in fact truncated by the formation of electron and hole puddles that screen the disorder potential, as is the case with narrow band gap semiconductors\cite{shklovskii_completely_1972, shklovskii_electronic_1984, rossi_inhomogenous_2011, skinner_why_2012} and two-dimensional Dirac systems like graphene\cite{shklovskii_simple_2007, galitski_statistics_2007} and topological insulators. \cite{beidenkopf_spatial_2011, skinner_effects_2013}  This disorder-induced puddling is shown schematically in Fig.\ \ref{fig:disorder-schematic}.  It is the purpose of this paper to calculate the typical size and density of these puddles, as well as the corresponding disorder potential amplitude, DOS, and conductivity.

The question of disorder effects in 3DDSs has acquired a particular relevance in recent months, following the experimental discovery of two different 3D Dirac materials \cite{borisenko_experimental_2013, neupane_discovery_2013, liu_discovery_2014, jeon_landau_2014} not long after their theoretical prediction.\cite{murakami_phase_2007, wan_topological_2011, burkov_weyl_2011, turner_beyond_2013, young_dirac_2012}  While the existence of a Dirac dispersion in these materials has been established, largely by photoemission experiments, it remains to be thoroughly understood how closely the Dirac point can be probed and to what extent its behavior is masked by disorder.  As shown below, the presence of Coulomb impurities has the effect of ``smearing" the Dirac point via the creation of electron and hole puddles, and this smearing typically occurs over tens of meV.

The structure and primary results of this paper are as follows.  
In Sec.\ \ref{sec:theory}, a self-consistent theory is developed to describe the disorder potential based on the Thomas-Fermi (TF) approximation.  Corresponding expressions are derived for the magnitude of the disorder potential [\eq{G0}], the DOS [\eq{nu0}], the size of electron/hole puddles [\eq{rs0}], and the concentration of electrons/holes in puddles [\eq{np}].  While the primary focus of this section is on the behavior near the Dirac point, results are also presented for the case when the chemical potential is away from the Dirac point.  In Sec.\ \ref{sec:conductivity}, the zero-temperature conductivity is discussed, and a result is presented for the conductivity [\eq{sigman}] and its minimum value [\eq{smin}].  Sec.\ \ref{sec:conclusion} concludes with a summary and some discussion of recent experiments.

\section{Self-consistent theory of disorder and screening}
\label{sec:theory}

This paper focuses on a model of disorder in which $N$ monovalent Coulomb impurities per unit volume are randomly distributed throughout the bulk of a 3DDS.  Such impurities create a random Coulomb potential, and this potential induces an electron/hole concentration $n(\rr)$, where $n > 0$ for electrons and $n < 0$ for holes.  The value of $n(\rr)$ at a given spatial coordinate $\rr$ is related to the self-consistent magnitude of the Coulomb potential $\phi(\rr)$.  In this paper, the primary tool for describing this relationship is the TF approximation:
\be 
E_f[n(\rr)] - e \phi(\rr) = \mu.
\label{eq:TF}
\ee 
Here, $E_f(n) = \hbar v k_f(n) \sgn(n)$ is the local Fermi energy, where $k_f$ is the Fermi wave vector and $\mu$ is the chemical potential of the 3DDS measured relative to the Dirac point.  Assuming, generically, that the Dirac point has a degeneracy $g$ (which for Weyl semimetals is equal to the number of degenerate Dirac points), the Fermi wave vector is $k_f = (6 \pi^2 |n|/g)^{1/3}$, so that $E_f(n) = (6 \pi^2/g)^{1/3} \hbar v |n|^{1/3} \sgn(n)$.

If the chemical potential $\mu$ is large enough in absolute value that $|e \phi| \ll |\mu|$, one can think that the electron density is relatively uniform spatially, and the corresponding electron DOS $\nu = dn/dE_f = g E_f^2/(2 \pi^2 \hbar^3 v^3) \simeq g \mu^2/(2 \pi^2 \hbar^3 v^3)$ is also uniform.  In this case, one can straightforwardly define a TF screening radius
\be 
r_s = \sqrt{\frac{\kappa}{4 \pi e^2 \nu}} = \sqrt{\frac{\pi}{2 \alpha g}} k_f^{-1}.
\label{eq:rs}
\ee 
Here, $\alpha = e^2/\kappa \hbar v$ is the effective fine structure constant.  The TF approximation is valid in cases where the Fermi wavelength $\sim k_f^{-1}$ is much shorter than the typical scale over which the potential varies, $r_s$.  As can be seen in \eq{rs}, this corresponds to $\alpha g \ll 1$.  For comparison, the 3DDS \CA has $\hbar v \approx 5$--$10$ eV$\cdot$\AA,\cite{neupane_discovery_2013, borisenko_experimental_2013, jeon_landau_2014} $\kappa \approx 36$, and $g = 2$, so that $\alpha g \approx 0.08$--$0.16$ and this approximation is justified.  As shown below, the same criterion $\alpha g \ll 1$ justifies the use of the TF approximation for the case of $\mu = 0$.

If $|\mu|$ is not large, so that the 3DDS is close to the Dirac point, then one cannot consider the electron density to be uniform and the typical screening radius $r_s$ must be found self-consistently.  In particular, one can assume that the disorder potential is screened with some unknown screening radius $r_s$ and then calculate analytically the corresponding magnitude of the disorder potential and the resulting average density of states $\avg{ \nu}$.  Inserting the result for $\nu$ into \eq{rs}, one arrives at a self-consistent relationship for $r_s$, which can be solved to give a result for $r_s$, $\avg{\nu}$, and the magnitude of the disorder potential.\cite{stern_low-temperature_1974}  This procedure is carried out explicitly in the remainder of the present section.

In a medium with screening radius $r_s$, the screened potential produced by a single impurity with charge $\pm e$ is the Yukawa-like potential\cite{mahan_many-particle_1990}
\be 
\phi_1(r) = \pm \frac{e}{\kappa r} \exp[-r/r_s].
\ee 
If one assumes that impurity positions are uncorrelated, then the mean squared value of the electron potential energy, $\Gamma^2$, can be found by integrating the square of the potential created by a single impurity, $(e \phi_1)^2$, over all possible impurity positions.  This gives
\be
\Gamma^2 = \int \left(e \phi_1(r)\right)^2 N d^3r 
 = 2 \pi e^4 N r_s/\kappa. 
\label{eq:G2rs}
\ee
For cases where $r_s \gg N^{-1/3}$ (justified below), the potential at each point in space is the sum of the potentials produced by many independently-located impurities.  By the central limit theorem, then, one can assume that the distribution of values of the potential across the system is Gaussian with variance $\Gamma^2/e^2$.  Within the TF approximation, the value of the DOS at a point with potential $\phi$ is $\nu(\phi) = (g/2 \pi^2 \hbar^3 v^3)(\mu + e \phi)^2$, so that the spatially-averaged DOS is
\begin{eqnarray} 
\avg{\nu} & = & \int_{-\infty}^{\infty} \nu(\phi) \frac{\exp\left[ -e^2\phi^2/2 \Gamma^2 \right]}{\sqrt{2 \pi \Gamma^2/e^2}}  \, d\phi \nonumber \\
& = & \frac{g}{2 \pi^2 \hbar^3 v^3}(\Gamma^2 + \mu^2).
\label{eq:avgnu}
\end{eqnarray}
Inserting this expression for $\nu$ in \eq{rs} and plugging the resulting expression for $r_s$ into \eq{G2rs} gives the following self-consistent expression for the amplitude $\Gamma$ of the disorder potential:
\be 
\Gamma^4 (\Gamma^2 + \mu^2) = \frac{2 \pi^3}{g \alpha^3} \left( \frac{e^2 N^{1/3}}{\kappa} \right)^6.
\label{eq:Gself}
\ee 

Equation (\ref{eq:Gself}) can be solved for generic values of the chemical potential $\mu$, but it is worth considering specifically the cases of $\mu = 0$ (when the 3DDS is at the Dirac point) and large $|\mu|$.  When $\mu = 0$, \eq{Gself} gives
\be 
\Gamma \equiv \Gamma_0 = \left(\frac{2 \pi^3}{g \alpha^3}\right)^{1/6} \frac{e^2 N^{1/3}}{\kappa}.
\label{eq:G0}
\ee 
It is perhaps worth noting that this value for the disorder potential amplitude is smaller than the corresponding result for the surface of a disordered 3D topological insulator\cite{skinner_theory_2013} by a factor $\sim \alpha^{1/6}$.  This smaller disorder for 3DDSs is a consequence of stronger screening in three dimensions.

Inserting the value of $\Gamma_0$ from \eq{G0} into \eq{avgnu} gives the corresponding DOS at $\mu = 0$:
\be 
\nu_0 = \left( \frac{\alpha^3 g^2}{4 \pi^3} \right)^{1/3} \frac{N^{2/3}}{\hbar v}.
\label{eq:nu0}
\ee 
In this case, the screening radius $r_s$, which is generically equal to the correlation length of the disorder potential, defines the typical size of electron and hole puddles (as illustrated in Fig.\ \ref{fig:disorder-schematic}).  By \eq{rs}, its value is given by 
\be 
r_s = \left( \frac{1}{4 g \alpha^3} \right)^{1/3} N^{-1/3}.
\label{eq:rs0}
\ee 
The typical concentration of electrons/holes in puddles $n_p$ is found by equating $\Gamma_0$ with $E_f(n_p)$, which gives
\be 
n_p = \sqrt{\frac{g \alpha^3}{18 \pi}} N
\label{eq:np},
\ee 
so that the corresponding number of electrons/holes per puddle is
\be 
M_p \approx \frac{4 \pi}{3} r_s^3 n_p = \sqrt{\frac{\pi/162}{g \alpha^3}}.
\label{eq:Mp}
\ee 
When $g \alpha \ll 1$, there are many electrons per puddle: $M_p \gg 1$.  Intriguingly, this value for $M_p$ is independent of the impurity concentration, so that the number of electrons per puddle is independent of the details of the disorder.  This universality is reminiscent of the problem of a single supercritical nucleus in a 3DDS, where the maximum observable ``nuclear charge" also obtains a universal value $\sim 1/\alpha^{3/2}$.\cite{kolomeisky_fermion_2013}

Notice also that at $g \alpha \ll 1$, the correlation length of the potential $r_s$ is much longer than the typical Fermi wavelength $k_f(n_p)^{-1} \sim N^{-1/3} g^{1/6}/\alpha^{1/2}$, so that the TF approximation is justified.  This same condition also guarantees $r_s \gg N^{-1/3}$, which validates the assumption of a Gaussian-distributed potential.

It is worth noting that Eqs.\ (\ref{eq:G0})--(\ref{eq:Mp}) can be derived qualitatively using the following very simple argument (which for simplicity uses $g \sim 1$).  Consider a volume of size $\sim r_s$ within the 3DDS; this volume is effectively a single electron/hole puddle.  Those impurities within the volume can be said to contribute to the potential within it, while others are effectively screened out.  The net charge of impurities in the volume is $Q \sim e\sqrt{N r_s^3}$ (with a random sign), and this impurity charge is compensated by the charge of electrons/holes, which have total number $M_p \sim n_p r_s^3$.  Equating $M_p$ with $Q/e$ gives $n_p^2 \sim N/r_s^3$.  Now one can note that the typical kinetic energy of electrons within the volume, $\sim \hbar v n_p^{1/3}$, must be similar in magnitude to the typical Coulomb energy $\sim Q e/\kappa r_s$.  This equality gives $n_p^2 \sim \alpha^6 N^3 r_s^3$.  Combining the two equations for $n_p$ gives $r_s \sim N^{-1/3}/\alpha$, as in \eq{rs0}, and the other relevant quantities can be found by substitution.

As the chemical potential $\mu$ is moved away from the Dirac point, the magnitude of the disorder potential decreases, as dictated by \eq{Gself}, and correspondingly the screening radius $r_s$ shrinks.  At $|\mu| \gg \Gamma_0$, puddles of electrons (for $\mu < 0$) or holes (for $\mu > 0$) dry up, and the system is well-described by linear screening with a spatially-uniform DOS.  In this case the disorder potential magnitude becomes
\be 
\Gamma \simeq \left( \frac{2 \pi^3}{g \alpha^3} \right)^{1/4} \left( \frac{e^2 N^{1/3}/\kappa}{|\mu|} \right)^{1/2} \frac{e^2 N^{1/3}}{\kappa}.
\label{eq:Glarge}
\ee 
The corresponding DOS approaches that of the non-disordered system,
\be 
\nu \simeq \frac{g}{2 \pi^2} \frac{\mu^2}{ (\hbar v)^3}
\label{eq:nularge}
\ee 
and the correlation length of the disorder potential is 
\be 
r_s \simeq \sqrt{\frac{\pi}{2 \alpha g} } \frac{\hbar v}{|\mu|}.
\label{eq:rslarge}
\ee 

\begin{figure}[htb!]
\centering
\includegraphics[width=0.45 \textwidth]{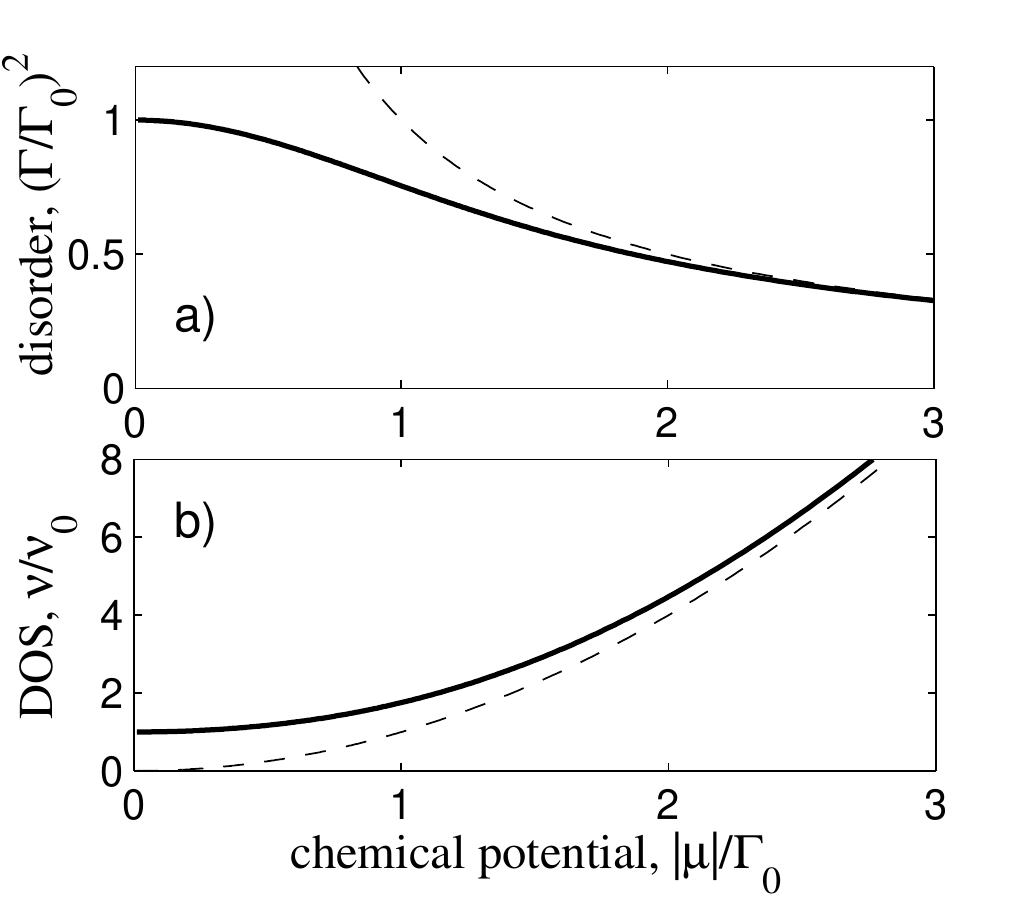}
\caption{a) The variance in the disorder potential as a function of the chemical potential $\mu$.  The dashed line corresponds to the linear screening regime of \eq{Glarge}.  b) The spatially-averaged density of states as a function of the chemical potential.  The dashed line corresponds to \eq{nularge}, which is the DOS for a non-disordered system.  $\Gamma_0$ and $\nu_0$ are given by Eqs.\ (\ref{eq:G0}) and (\ref{eq:nu0}), respectively. }
\label{fig:Gamma-DOS}
\end{figure}

Equations (\ref{eq:G0})--(\ref{eq:Mp}) and (\ref{eq:Glarge})--(\ref{eq:rslarge}) describe the system in the limits of $\mu = 0$ and $|\mu| \gg \Gamma_0$, respectively.  The crossover between these two regimes can be described by evaluating Eqs.\ (\ref{eq:avgnu}) and (\ref{eq:Gself}).  The result of this process is shown in Fig.\ \ref{fig:Gamma-DOS}, where the variance of the disorder potential and the DOS are plotted as a function of the chemical potential $\mu$.  As one can see, these self-consistent equations predict a smooth, monotonic crossover from the puddle-dominated $\mu = 0$ result to the linear screening regime at large $\mu$.  It is worth noting, however, that $\Gamma$ may in fact exhibit weakly nonmonotonic behavior as a function of $\mu$, achieving a weak maximum at $|\mu|/\Gamma_0 \sim 1$.  Such behavior is predicted theoretically for topological insulators,\cite{skinner_theory_2013} and arises because at $|\mu|/\Gamma_0 \sim 1$ the distribution of values of the Coulomb potential becomes skewed toward those values that bring the system locally closer to the Dirac point, where screening is poorer.  

\section{Conductivity}
\label{sec:conductivity}

The previous section discussed the disorder potential produced by random Coulomb impurities using a self-consistent theory of screening.  In this section I briefly discuss the implications of this screening for the low-temperature conductivity.

In situations where the mean free path $\ell$ for electron scattering is relatively large, $k_f \ell \gg 1$, the electron transport is well-described by the Boltzmann equation.  In particular, the momentum relaxation time $\tau$ satisfies
\be 
\frac{\hbar}{\tau} = \pi N \nu \int_0^\infty d\theta \sin \theta \left| \pt_1(q) \right|^2 (1 - \cos \theta) \frac{1 + \cos \theta}{2}
\label{eq:Boltzmann}
\ee 
(see, e.g., Ref.\ \onlinecite{burkov_topological_2011}).  Here, $q = 2 k_f \sin(\theta/2)$ is the momentum change resulting from scattering by an angle $\theta$ and $\pt_1(q) = 4 \pi e^2/[\kappa (q^2 + r_s^{-2})]$ is the Fourier transform of $\phi_1(r)$ evaluated at wave vector $q$.  The final factor of $(1 + \cos \theta)/2$ in \eq{Boltzmann} arises when backscattering is suppressed as a consequence of the spin texture at the Dirac point, as in Weyl semimetals; omitting this factor does not change the results to leading order in $k_f r_s \gg 1$.  Evaluating the integral in \eq{Boltzmann} gives
\be 
\frac{\hbar}{\tau} \simeq \frac{8 \pi^3 N \nu e^4}{\kappa^2 k_f^4} \ln(2 k_f r_s)
\nonumber
\ee 
assuming $k_f r_s \gg 1$, which is a condition of validity for the screened potential used here, and, again, corresponds to $\alpha g \ll 1$.

One can arrive at an expression for the conductivity $\sigma$ by combining the expression for $\tau$ with the Einstein relation for conductivity, $\sigma = e^2 \nu (v^2 \tau/3)$.\cite{abrikosov_alexei_a_fundamentals_1988}  If one assumes a relatively large and uniform electron density $n$ (i.e., a large chemical potential $|\mu| \gg \Gamma_0$), then $k_f \simeq (6 \pi^2 n/g)^{1/3}$ and
\be 
\sigma = \left( \frac{3}{4 \pi} \right)^{1/3} \frac{1}{\alpha^2 g^{4/3} \ln(2 \pi / \alpha g)} \frac{n}{N} \frac{e^2 n^{1/3}}{\hbar}.
\label{eq:sigman}
\ee 
As the electron density is reduced (the chemical potential is brought closer to the Dirac point), the number of carriers is reduced and the conductivity declines.  At $\mu = 0$, the conductivity achieves a minimum whose value is determined by the concentration of electrons and holes in puddles.  The value of this conductivity minimum, $\sigma_\text{min}$, can be estimated by inserting $n \sim n_p$ into \eq{sigman}, which gives
\be 
\sigma_\text{min} \approx \frac{1}{6 \pi (2g^2)^{1/3} \ln (2 \pi/\alpha g)} \frac{e^2 N^{1/3}}{\hbar}
\label{eq:smin}.
\ee 

Finally, one can check that the derived expressions indeed correspond to the Boltzmann semiclassical limit $k_f \ell \gg 1$.  As expected, the value of $\tau$ corresponding to the minimum conductivity, where $n \sim n_p$, gives $k_f \ell = k_f v \tau \sim g^{2/3}/\alpha g \gg 1$.  Thus, $k_f \ell \gg 1$, provided that $\alpha g \ll 1$.  At larger $n$ the mean free path only increases, so that the Boltzmann equation is a good description everywhere.

\section{Concluding remarks}
\label{sec:conclusion}

This paper has presented a simple picture of self-consistent screening of Coulomb impurities in 3DDSs through the formation of electron and hole puddles.  Such effects are manifestly not perturbative near the Dirac point, regardless of the impurity concentration, and they have a prominent effect on both the observed DOS and the conductivity.  The DOS, for example, vanishes  only as the $2/3$ power of the impurity concentration, which suggests that a true Dirac semimetal phase with vanishing DOS may remain frustratingly elusive experimentally.

While the experimental study of 3DDSs is still very young, one can get a sense of the typical scales of the disorder potential by using parameters for the recently-discovered bulk Dirac material \CA, which has $\alpha \approx 0.04$ and $g = 2$.  Thus far, experimentally studied samples are $n$-type, apparently resulting from uncontrolled doping by As vacancies.\cite{jeon_landau_2014, neupane_discovery_2013, borisenko_experimental_2013} For example, Ref.\ \onlinecite{jeon_landau_2014} reports relatively large $\mu \approx 200$\,meV, with a corresponding carrier concentration $n \sim 2 \times 10^{18}$\,cm$^{-3}$; one can expect that the concentration of donor impurities $N$ is similar in magnitude.  Inserting these parameters into Eqs.\ (\ref{eq:Glarge}) and (\ref{eq:rslarge}) gives an estimated disorder potential of $\Gamma \approx 20$\,meV and a screening radius $r_s \approx 20$\,nm.  The latter seems consistent with the scale of disorder fluctuations seen by scanning tunneling microscopy measurements.\cite{jeon_landau_2014}  By Equation (\ref{eq:sigman}), this level of disorder corresponds to a mobility $\sigma/(e n) \sim 30\,000$\,cm$^2$/Vs, which also closely matches the value seen in experiment.\cite{neupane_discovery_2013}

Future efforts to bring the bulk chemical potential of 3DDSs to the Dirac point will presumably require compensation of donors by acceptors.  By Eqs.\ (\ref{eq:np}) and (\ref{eq:Mp}), the resulting disorder landscape can be expected to have a typical concentration $n_p \sim 10^{-3}N \sim 10^{-15}$\,cm$^{-3}$ of electrons in puddles and $\sim 10$ electrons/holes per puddle, with a disorder potential of magnitude $\Gamma_0 \sim 45$\,meV, assuming the impurity concentration $N$ remains of order $10^{18}$\,cm$^{-3}$.  Equation (\ref{eq:smin}) suggests a corresponding minimum conductivity $\sigma_\text{min} \sim 1$\,S/cm. 

Finally, one can note that existing 3DDS materials seem to have anisotropic Dirac cones, with a Dirac velocity in one particular direction, $v_z$, that is as much as ten times smaller than the velocity in the transverse directions, $v_\perp$.\cite{neupane_discovery_2013}  This anisotropy can be accounted for at the level of the present theory by substituting for $v$ the geometric mean velocity $(v_\perp^2 v_z)^{1/3}$, so that the fine structure constant $\alpha \propto 1/v$ is also modified.

\acknowledgments

I am grateful to R. Nandkishore, S. Gopalakrishnan, J. C. W. Song, B. I. Shklovskii, and E. B. Kolomeisky for helpful discussions and comments.  
Work at Argonne National Laboratory was supported by the U.S. Department of Energy, Office of Basic Energy Sciences under contract no. DE-AC02-06CH11357.

\bibliography{Dirac3D}

\begin{thebibliography}{30}%
\makeatletter
\providecommand \@ifxundefined [1]{%
 \@ifx{#1\undefined}
}%
\providecommand \@ifnum [1]{%
 \ifnum #1\expandafter \@firstoftwo
 \else \expandafter \@secondoftwo
 \fi
}%
\providecommand \@ifx [1]{%
 \ifx #1\expandafter \@firstoftwo
 \else \expandafter \@secondoftwo
 \fi
}%
\providecommand \natexlab [1]{#1}%
\providecommand \enquote  [1]{``#1''}%
\providecommand \bibnamefont  [1]{#1}%
\providecommand \bibfnamefont [1]{#1}%
\providecommand \citenamefont [1]{#1}%
\providecommand \href@noop [0]{\@secondoftwo}%
\providecommand \href [0]{\begingroup \@sanitize@url \@href}%
\providecommand \@href[1]{\@@startlink{#1}\@@href}%
\providecommand \@@href[1]{\endgroup#1\@@endlink}%
\providecommand \@sanitize@url [0]{\catcode `\\12\catcode `\$12\catcode
  `\&12\catcode `\#12\catcode `\^12\catcode `\_12\catcode `\%12\relax}%
\providecommand \@@startlink[1]{}%
\providecommand \@@endlink[0]{}%
\providecommand \url  [0]{\begingroup\@sanitize@url \@url }%
\providecommand \@url [1]{\endgroup\@href {#1}{\urlprefix }}%
\providecommand \urlprefix  [0]{URL }%
\providecommand \Eprint [0]{\href }%
\providecommand \doibase [0]{http://dx.doi.org/}%
\providecommand \selectlanguage [0]{\@gobble}%
\providecommand \bibinfo  [0]{\@secondoftwo}%
\providecommand \bibfield  [0]{\@secondoftwo}%
\providecommand \translation [1]{[#1]}%
\providecommand \BibitemOpen [0]{}%
\providecommand \bibitemStop [0]{}%
\providecommand \bibitemNoStop [0]{.\EOS\space}%
\providecommand \EOS [0]{\spacefactor3000\relax}%
\providecommand \BibitemShut  [1]{\csname bibitem#1\endcsname}%
\let\auto@bib@innerbib\@empty
\bibitem [{\citenamefont {Shklovskii}\ and\ \citenamefont
  {Efros}(1984)}]{shklovskii_electronic_1984}%
  \BibitemOpen
  \bibfield  {author} {\bibinfo {author} {\bibfnamefont {B.~I.}\ \bibnamefont
  {Shklovskii}}\ and\ \bibinfo {author} {\bibfnamefont {A.~L.}\ \bibnamefont
  {Efros}},\ }\href@noop {} {\emph {\bibinfo {title} {Electronic Properties of
  Doped Semiconductors}}}\ (\bibinfo  {publisher} {Springer-Verlag},\ \bibinfo
  {address} {New York},\ \bibinfo {year} {1984})\BibitemShut {NoStop}%
\bibitem [{\citenamefont {Goswami}\ and\ \citenamefont
  {Chakravarty}(2011)}]{goswami_quantum_2011}%
  \BibitemOpen
  \bibfield  {author} {\bibinfo {author} {\bibfnamefont {P.}~\bibnamefont
  {Goswami}}\ and\ \bibinfo {author} {\bibfnamefont {S.}~\bibnamefont
  {Chakravarty}},\ }\href {\doibase 10.1103/PhysRevLett.107.196803} {\bibfield
  {journal} {\bibinfo  {journal} {Physical Review Letters}\ }\textbf {\bibinfo
  {volume} {107}},\ \bibinfo {pages} {196803} (\bibinfo {year}
  {2011})}\BibitemShut {NoStop}%
\bibitem [{\citenamefont {Hosur}\ \emph {et~al.}(2012)\citenamefont {Hosur},
  \citenamefont {Parameswaran},\ and\ \citenamefont
  {Vishwanath}}]{hosur_charge_2012}%
  \BibitemOpen
  \bibfield  {author} {\bibinfo {author} {\bibfnamefont {P.}~\bibnamefont
  {Hosur}}, \bibinfo {author} {\bibfnamefont {S.~A.}\ \bibnamefont
  {Parameswaran}}, \ and\ \bibinfo {author} {\bibfnamefont {A.}~\bibnamefont
  {Vishwanath}},\ }\href {\doibase 10.1103/PhysRevLett.108.046602} {\bibfield
  {journal} {\bibinfo  {journal} {Physical Review Letters}\ }\textbf {\bibinfo
  {volume} {108}},\ \bibinfo {pages} {046602} (\bibinfo {year}
  {2012})}\BibitemShut {NoStop}%
\bibitem [{\citenamefont {Kobayashi}\ \emph {et~al.}(2014)\citenamefont
  {Kobayashi}, \citenamefont {Ohtsuki}, \citenamefont {Imura},\ and\
  \citenamefont {Herbut}}]{kobayashi_density_2014}%
  \BibitemOpen
  \bibfield  {author} {\bibinfo {author} {\bibfnamefont {K.}~\bibnamefont
  {Kobayashi}}, \bibinfo {author} {\bibfnamefont {T.}~\bibnamefont {Ohtsuki}},
  \bibinfo {author} {\bibfnamefont {K.-I.}\ \bibnamefont {Imura}}, \ and\
  \bibinfo {author} {\bibfnamefont {I.~F.}\ \bibnamefont {Herbut}},\ }\href
  {\doibase 10.1103/PhysRevLett.112.016402} {\bibfield  {journal} {\bibinfo
  {journal} {Physical Review Letters}\ }\textbf {\bibinfo {volume} {112}},\
  \bibinfo {pages} {016402} (\bibinfo {year} {2014})}\BibitemShut {NoStop}%
\bibitem [{\citenamefont {Syzranov}\ \emph {et~al.}(2014)\citenamefont
  {Syzranov}, \citenamefont {Radzihovsky},\ and\ \citenamefont
  {Gurarie}}]{syzranov_critical_2014}%
  \BibitemOpen
  \bibfield  {author} {\bibinfo {author} {\bibfnamefont {S.~V.}\ \bibnamefont
  {Syzranov}}, \bibinfo {author} {\bibfnamefont {L.}~\bibnamefont
  {Radzihovsky}}, \ and\ \bibinfo {author} {\bibfnamefont {V.}~\bibnamefont
  {Gurarie}},\ }\href {http://arxiv.org/abs/1402.3737} {\bibfield  {journal}
  {\bibinfo  {journal} {{arXiv:1402.3737} [cond-mat]}\ } (\bibinfo {year}
  {2014})}\BibitemShut {NoStop}%
\bibitem [{\citenamefont {Sbierski}\ \emph {et~al.}(2014)\citenamefont
  {Sbierski}, \citenamefont {Pohl}, \citenamefont {Bergholtz},\ and\
  \citenamefont {Brouwer}}]{sbierski_quantum_2014}%
  \BibitemOpen
  \bibfield  {author} {\bibinfo {author} {\bibfnamefont {B.}~\bibnamefont
  {Sbierski}}, \bibinfo {author} {\bibfnamefont {G.}~\bibnamefont {Pohl}},
  \bibinfo {author} {\bibfnamefont {E.~J.}\ \bibnamefont {Bergholtz}}, \ and\
  \bibinfo {author} {\bibfnamefont {P.~W.}\ \bibnamefont {Brouwer}},\ }\href
  {http://arxiv.org/abs/1402.6653} {\bibfield  {journal} {\bibinfo  {journal}
  {{arXiv:1402.6653} [cond-mat]}\ } (\bibinfo {year} {2014})}\BibitemShut
  {NoStop}%
\bibitem [{\citenamefont {Ominato}\ and\ \citenamefont
  {Koshino}(2014)}]{ominato_quantum_2014}%
  \BibitemOpen
  \bibfield  {author} {\bibinfo {author} {\bibfnamefont {Y.}~\bibnamefont
  {Ominato}}\ and\ \bibinfo {author} {\bibfnamefont {M.}~\bibnamefont
  {Koshino}},\ }\href {\doibase 10.1103/PhysRevB.89.054202} {\bibfield
  {journal} {\bibinfo  {journal} {Physical Review B}\ }\textbf {\bibinfo
  {volume} {89}},\ \bibinfo {pages} {054202} (\bibinfo {year}
  {2014})}\BibitemShut {NoStop}%
\bibitem [{\citenamefont {Nandkishore}\ \emph {et~al.}(2014)\citenamefont
  {Nandkishore}, \citenamefont {Huse},\ and\ \citenamefont
  {Sondhi}}]{nandkishore_rare_2014}%
  \BibitemOpen
  \bibfield  {author} {\bibinfo {author} {\bibfnamefont {R.}~\bibnamefont
  {Nandkishore}}, \bibinfo {author} {\bibfnamefont {D.~A.}\ \bibnamefont
  {Huse}}, \ and\ \bibinfo {author} {\bibfnamefont {S.~L.}\ \bibnamefont
  {Sondhi}},\ }\href {http://arxiv.org/abs/1405.2336} {\bibfield  {journal}
  {\bibinfo  {journal} {{arXiv:1405.2336} [cond-mat]}\ } (\bibinfo {year}
  {2014})}\BibitemShut {NoStop}%
\bibitem [{\citenamefont {Shklovskii}\ and\ \citenamefont
  {Efros}(1972)}]{shklovskii_completely_1972}%
  \BibitemOpen
  \bibfield  {author} {\bibinfo {author} {\bibfnamefont {B.~I.}\ \bibnamefont
  {Shklovskii}}\ and\ \bibinfo {author} {\bibfnamefont {A.~L.}\ \bibnamefont
  {Efros}},\ }\href@noop {} {\bibfield  {journal} {\bibinfo  {journal} {Sov.
  Phys.-{JETP}}\ }\textbf {\bibinfo {volume} {35}},\ \bibinfo {pages} {610}
  (\bibinfo {year} {1972})}\BibitemShut {NoStop}%
\bibitem [{\citenamefont {Rossi}\ and\ \citenamefont
  {Das~Sarma}(2011)}]{rossi_inhomogenous_2011}%
  \BibitemOpen
  \bibfield  {author} {\bibinfo {author} {\bibfnamefont {E.}~\bibnamefont
  {Rossi}}\ and\ \bibinfo {author} {\bibfnamefont {S.}~\bibnamefont
  {Das~Sarma}},\ }\href {\doibase 10.1103/PhysRevLett.107.155502} {\bibfield
  {journal} {\bibinfo  {journal} {Physical Review Letters}\ }\textbf {\bibinfo
  {volume} {107}},\ \bibinfo {pages} {155502} (\bibinfo {year}
  {2011})}\BibitemShut {NoStop}%
\bibitem [{\citenamefont {Skinner}\ \emph {et~al.}(2012)\citenamefont
  {Skinner}, \citenamefont {Chen},\ and\ \citenamefont
  {Shklovskii}}]{skinner_why_2012}%
  \BibitemOpen
  \bibfield  {author} {\bibinfo {author} {\bibfnamefont {B.}~\bibnamefont
  {Skinner}}, \bibinfo {author} {\bibfnamefont {T.}~\bibnamefont {Chen}}, \
  and\ \bibinfo {author} {\bibfnamefont {B.~I.}\ \bibnamefont {Shklovskii}},\
  }\href {\doibase 10.1103/PhysRevLett.109.176801} {\bibfield  {journal}
  {\bibinfo  {journal} {Physical Review Letters}\ }\textbf {\bibinfo {volume}
  {109}},\ \bibinfo {pages} {176801} (\bibinfo {year} {2012})}\BibitemShut
  {NoStop}%
\bibitem [{\citenamefont {Shklovskii}(2007)}]{shklovskii_simple_2007}%
  \BibitemOpen
  \bibfield  {author} {\bibinfo {author} {\bibfnamefont {B.~I.}\ \bibnamefont
  {Shklovskii}},\ }\href {\doibase 10.1103/PhysRevB.76.233411} {\bibfield
  {journal} {\bibinfo  {journal} {Phys. Rev. B}\ }\textbf {\bibinfo {volume}
  {76}},\ \bibinfo {pages} {233411} (\bibinfo {year} {2007})}\BibitemShut
  {NoStop}%
\bibitem [{\citenamefont {Galitski}\ \emph {et~al.}(2007)\citenamefont
  {Galitski}, \citenamefont {Adam},\ and\ \citenamefont
  {Das~Sarma}}]{galitski_statistics_2007}%
  \BibitemOpen
  \bibfield  {author} {\bibinfo {author} {\bibfnamefont {V.~M.}\ \bibnamefont
  {Galitski}}, \bibinfo {author} {\bibfnamefont {S.}~\bibnamefont {Adam}}, \
  and\ \bibinfo {author} {\bibfnamefont {S.}~\bibnamefont {Das~Sarma}},\ }\href
  {\doibase 10.1103/PhysRevB.76.245405} {\bibfield  {journal} {\bibinfo
  {journal} {Physical Review B}\ }\textbf {\bibinfo {volume} {76}},\ \bibinfo
  {pages} {245405} (\bibinfo {year} {2007})}\BibitemShut {NoStop}%
\bibitem [{\citenamefont {Beidenkopf}\ \emph {et~al.}(2011)\citenamefont
  {Beidenkopf}, \citenamefont {Roushan}, \citenamefont {Seo}, \citenamefont
  {Gorman}, \citenamefont {Drozdov}, \citenamefont {Hor}, \citenamefont
  {Cava},\ and\ \citenamefont {Yazdani}}]{beidenkopf_spatial_2011}%
  \BibitemOpen
  \bibfield  {author} {\bibinfo {author} {\bibfnamefont {H.}~\bibnamefont
  {Beidenkopf}}, \bibinfo {author} {\bibfnamefont {P.}~\bibnamefont {Roushan}},
  \bibinfo {author} {\bibfnamefont {J.}~\bibnamefont {Seo}}, \bibinfo {author}
  {\bibfnamefont {L.}~\bibnamefont {Gorman}}, \bibinfo {author} {\bibfnamefont
  {I.}~\bibnamefont {Drozdov}}, \bibinfo {author} {\bibfnamefont {Y.~S.}\
  \bibnamefont {Hor}}, \bibinfo {author} {\bibfnamefont {R.~J.}\ \bibnamefont
  {Cava}}, \ and\ \bibinfo {author} {\bibfnamefont {A.}~\bibnamefont
  {Yazdani}},\ }\href {\doibase 10.1038/nphys2108} {\bibfield  {journal}
  {\bibinfo  {journal} {Nature Physics}\ }\textbf {\bibinfo {volume} {7}},\
  \bibinfo {pages} {939} (\bibinfo {year} {2011})}\BibitemShut {NoStop}%
\bibitem [{\citenamefont {Skinner}\ \emph {et~al.}(2013)\citenamefont
  {Skinner}, \citenamefont {Chen},\ and\ \citenamefont
  {Shklovskii}}]{skinner_effects_2013}%
  \BibitemOpen
  \bibfield  {author} {\bibinfo {author} {\bibfnamefont {B.}~\bibnamefont
  {Skinner}}, \bibinfo {author} {\bibfnamefont {T.}~\bibnamefont {Chen}}, \
  and\ \bibinfo {author} {\bibfnamefont {B.~I.}\ \bibnamefont {Shklovskii}},\
  }\href {\doibase 10.1134/S1063776113110150} {\bibfield  {journal} {\bibinfo
  {journal} {Journal of Experimental and Theoretical Physics}\ }\textbf
  {\bibinfo {volume} {117}},\ \bibinfo {pages} {579} (\bibinfo {year}
  {2013})}\BibitemShut {NoStop}%
\bibitem [{\citenamefont {Borisenko}\ \emph {et~al.}(2013)\citenamefont
  {Borisenko}, \citenamefont {Gibson}, \citenamefont {Evtushinsky},
  \citenamefont {Zabolotnyy}, \citenamefont {Buechner},\ and\ \citenamefont
  {Cava}}]{borisenko_experimental_2013}%
  \BibitemOpen
  \bibfield  {author} {\bibinfo {author} {\bibfnamefont {S.}~\bibnamefont
  {Borisenko}}, \bibinfo {author} {\bibfnamefont {Q.}~\bibnamefont {Gibson}},
  \bibinfo {author} {\bibfnamefont {D.}~\bibnamefont {Evtushinsky}}, \bibinfo
  {author} {\bibfnamefont {V.}~\bibnamefont {Zabolotnyy}}, \bibinfo {author}
  {\bibfnamefont {B.}~\bibnamefont {Buechner}}, \ and\ \bibinfo {author}
  {\bibfnamefont {R.~J.}\ \bibnamefont {Cava}},\ }\href
  {http://arxiv.org/abs/1309.7978} {\bibfield  {journal} {\bibinfo  {journal}
  {{arXiv:1309.7978} [cond-mat]}\ } (\bibinfo {year} {2013})}\BibitemShut
  {NoStop}%
\bibitem [{\citenamefont {Neupane}\ \emph {et~al.}(2013)\citenamefont
  {Neupane}, \citenamefont {Xu}, \citenamefont {Sankar}, \citenamefont
  {Alidoust}, \citenamefont {Bian}, \citenamefont {Liu}, \citenamefont
  {Belopolski}, \citenamefont {Chang}, \citenamefont {Jeng}, \citenamefont
  {Lin}, \citenamefont {Bansil}, \citenamefont {Chou},\ and\ \citenamefont
  {Hasan}}]{neupane_discovery_2013}%
  \BibitemOpen
  \bibfield  {author} {\bibinfo {author} {\bibfnamefont {M.}~\bibnamefont
  {Neupane}}, \bibinfo {author} {\bibfnamefont {S.}~\bibnamefont {Xu}},
  \bibinfo {author} {\bibfnamefont {R.}~\bibnamefont {Sankar}}, \bibinfo
  {author} {\bibfnamefont {N.}~\bibnamefont {Alidoust}}, \bibinfo {author}
  {\bibfnamefont {G.}~\bibnamefont {Bian}}, \bibinfo {author} {\bibfnamefont
  {C.}~\bibnamefont {Liu}}, \bibinfo {author} {\bibfnamefont {I.}~\bibnamefont
  {Belopolski}}, \bibinfo {author} {\bibfnamefont {T.-R.}\ \bibnamefont
  {Chang}}, \bibinfo {author} {\bibfnamefont {H.-T.}\ \bibnamefont {Jeng}},
  \bibinfo {author} {\bibfnamefont {H.}~\bibnamefont {Lin}}, \bibinfo {author}
  {\bibfnamefont {A.}~\bibnamefont {Bansil}}, \bibinfo {author} {\bibfnamefont
  {F.}~\bibnamefont {Chou}}, \ and\ \bibinfo {author} {\bibfnamefont {M.~Z.}\
  \bibnamefont {Hasan}},\ }\href {http://arxiv.org/abs/1309.7892} {\bibfield
  {journal} {\bibinfo  {journal} {{arXiv:1309.7892} [cond-mat]}\ } (\bibinfo
  {year} {2013})}\BibitemShut {NoStop}%
\bibitem [{\citenamefont {Liu}\ \emph {et~al.}(2014)\citenamefont {Liu},
  \citenamefont {Zhou}, \citenamefont {Zhang}, \citenamefont {Wang},
  \citenamefont {Weng}, \citenamefont {Prabhakaran}, \citenamefont {Mo},
  \citenamefont {Shen}, \citenamefont {Fang}, \citenamefont {Dai},
  \citenamefont {Hussain},\ and\ \citenamefont {Chen}}]{liu_discovery_2014}%
  \BibitemOpen
  \bibfield  {author} {\bibinfo {author} {\bibfnamefont {Z.~K.}\ \bibnamefont
  {Liu}}, \bibinfo {author} {\bibfnamefont {B.}~\bibnamefont {Zhou}}, \bibinfo
  {author} {\bibfnamefont {Y.}~\bibnamefont {Zhang}}, \bibinfo {author}
  {\bibfnamefont {Z.~J.}\ \bibnamefont {Wang}}, \bibinfo {author}
  {\bibfnamefont {H.~M.}\ \bibnamefont {Weng}}, \bibinfo {author}
  {\bibfnamefont {D.}~\bibnamefont {Prabhakaran}}, \bibinfo {author}
  {\bibfnamefont {S.-K.}\ \bibnamefont {Mo}}, \bibinfo {author} {\bibfnamefont
  {Z.~X.}\ \bibnamefont {Shen}}, \bibinfo {author} {\bibfnamefont
  {Z.}~\bibnamefont {Fang}}, \bibinfo {author} {\bibfnamefont {X.}~\bibnamefont
  {Dai}}, \bibinfo {author} {\bibfnamefont {Z.}~\bibnamefont {Hussain}}, \ and\
  \bibinfo {author} {\bibfnamefont {Y.~L.}\ \bibnamefont {Chen}},\ }\href
  {\doibase 10.1126/science.1245085} {\bibfield  {journal} {\bibinfo  {journal}
  {Science}\ }\textbf {\bibinfo {volume} {343}},\ \bibinfo {pages} {864}
  (\bibinfo {year} {2014})},\ \bibinfo {note} {{PMID:} 24436183}\BibitemShut
  {NoStop}%
\bibitem [{\citenamefont {Jeon}\ \emph {et~al.}(2014)\citenamefont {Jeon},
  \citenamefont {Zhou}, \citenamefont {Gyenis}, \citenamefont {Feldman},
  \citenamefont {Kimchi}, \citenamefont {Potter}, \citenamefont {Gibson},
  \citenamefont {Cava}, \citenamefont {Vishwanath},\ and\ \citenamefont
  {Yazdani}}]{jeon_landau_2014}%
  \BibitemOpen
  \bibfield  {author} {\bibinfo {author} {\bibfnamefont {S.}~\bibnamefont
  {Jeon}}, \bibinfo {author} {\bibfnamefont {B.~B.}\ \bibnamefont {Zhou}},
  \bibinfo {author} {\bibfnamefont {A.}~\bibnamefont {Gyenis}}, \bibinfo
  {author} {\bibfnamefont {B.~E.}\ \bibnamefont {Feldman}}, \bibinfo {author}
  {\bibfnamefont {I.}~\bibnamefont {Kimchi}}, \bibinfo {author} {\bibfnamefont
  {A.~C.}\ \bibnamefont {Potter}}, \bibinfo {author} {\bibfnamefont {Q.~D.}\
  \bibnamefont {Gibson}}, \bibinfo {author} {\bibfnamefont {R.~J.}\
  \bibnamefont {Cava}}, \bibinfo {author} {\bibfnamefont {A.}~\bibnamefont
  {Vishwanath}}, \ and\ \bibinfo {author} {\bibfnamefont {A.}~\bibnamefont
  {Yazdani}},\ }\href {http://arxiv.org/abs/1403.3446} {\bibfield  {journal}
  {\bibinfo  {journal} {{arXiv:1403.3446} [cond-mat]}\ } (\bibinfo {year}
  {2014})}\BibitemShut {NoStop}%
\bibitem [{\citenamefont {Murakami}(2007)}]{murakami_phase_2007}%
  \BibitemOpen
  \bibfield  {author} {\bibinfo {author} {\bibfnamefont {S.}~\bibnamefont
  {Murakami}},\ }\href {\doibase 10.1088/1367-2630/9/9/356} {\bibfield
  {journal} {\bibinfo  {journal} {New Journal of Physics}\ }\textbf {\bibinfo
  {volume} {9}},\ \bibinfo {pages} {356} (\bibinfo {year} {2007})}\BibitemShut
  {NoStop}%
\bibitem [{\citenamefont {Wan}\ \emph {et~al.}(2011)\citenamefont {Wan},
  \citenamefont {Turner}, \citenamefont {Vishwanath},\ and\ \citenamefont
  {Savrasov}}]{wan_topological_2011}%
  \BibitemOpen
  \bibfield  {author} {\bibinfo {author} {\bibfnamefont {X.}~\bibnamefont
  {Wan}}, \bibinfo {author} {\bibfnamefont {A.~M.}\ \bibnamefont {Turner}},
  \bibinfo {author} {\bibfnamefont {A.}~\bibnamefont {Vishwanath}}, \ and\
  \bibinfo {author} {\bibfnamefont {S.~Y.}\ \bibnamefont {Savrasov}},\ }\href
  {\doibase 10.1103/PhysRevB.83.205101} {\bibfield  {journal} {\bibinfo
  {journal} {Physical Review B}\ }\textbf {\bibinfo {volume} {83}},\ \bibinfo
  {pages} {205101} (\bibinfo {year} {2011})}\BibitemShut {NoStop}%
\bibitem [{\citenamefont {Burkov}\ and\ \citenamefont
  {Balents}(2011)}]{burkov_weyl_2011}%
  \BibitemOpen
  \bibfield  {author} {\bibinfo {author} {\bibfnamefont {A.~A.}\ \bibnamefont
  {Burkov}}\ and\ \bibinfo {author} {\bibfnamefont {L.}~\bibnamefont
  {Balents}},\ }\href {\doibase 10.1103/PhysRevLett.107.127205} {\bibfield
  {journal} {\bibinfo  {journal} {Physical Review Letters}\ }\textbf {\bibinfo
  {volume} {107}},\ \bibinfo {pages} {127205} (\bibinfo {year}
  {2011})}\BibitemShut {NoStop}%
\bibitem [{\citenamefont {Turner}\ and\ \citenamefont
  {Vishwanath}(2013)}]{turner_beyond_2013}%
  \BibitemOpen
  \bibfield  {author} {\bibinfo {author} {\bibfnamefont {A.~M.}\ \bibnamefont
  {Turner}}\ and\ \bibinfo {author} {\bibfnamefont {A.}~\bibnamefont
  {Vishwanath}},\ }\href {http://arxiv.org/abs/1301.0330} {\bibfield  {journal}
  {\bibinfo  {journal} {{arXiv:1301.0330} [cond-mat]}\ } (\bibinfo {year}
  {2013})}\BibitemShut {NoStop}%
\bibitem [{\citenamefont {Young}\ \emph {et~al.}(2012)\citenamefont {Young},
  \citenamefont {Zaheer}, \citenamefont {Teo}, \citenamefont {Kane},
  \citenamefont {Mele},\ and\ \citenamefont {Rappe}}]{young_dirac_2012}%
  \BibitemOpen
  \bibfield  {author} {\bibinfo {author} {\bibfnamefont {S.~M.}\ \bibnamefont
  {Young}}, \bibinfo {author} {\bibfnamefont {S.}~\bibnamefont {Zaheer}},
  \bibinfo {author} {\bibfnamefont {J.~C.~Y.}\ \bibnamefont {Teo}}, \bibinfo
  {author} {\bibfnamefont {C.~L.}\ \bibnamefont {Kane}}, \bibinfo {author}
  {\bibfnamefont {E.~J.}\ \bibnamefont {Mele}}, \ and\ \bibinfo {author}
  {\bibfnamefont {A.~M.}\ \bibnamefont {Rappe}},\ }\href {\doibase
  10.1103/PhysRevLett.108.140405} {\bibfield  {journal} {\bibinfo  {journal}
  {Physical Review Letters}\ }\textbf {\bibinfo {volume} {108}},\ \bibinfo
  {pages} {140405} (\bibinfo {year} {2012})}\BibitemShut {NoStop}%
\bibitem [{\citenamefont {Stern}(1974)}]{stern_low-temperature_1974}%
  \BibitemOpen
  \bibfield  {author} {\bibinfo {author} {\bibfnamefont {F.}~\bibnamefont
  {Stern}},\ }\href {\doibase 10.1103/PhysRevB.9.4597} {\bibfield  {journal}
  {\bibinfo  {journal} {Phys. Rev. B}\ }\textbf {\bibinfo {volume} {9}},\
  \bibinfo {pages} {4597–4598} (\bibinfo {year} {1974})}\BibitemShut
  {NoStop}%
\bibitem [{\citenamefont {Mahan}(1990)}]{mahan_many-particle_1990}%
  \BibitemOpen
  \bibfield  {author} {\bibinfo {author} {\bibfnamefont {G.~D.}\ \bibnamefont
  {Mahan}},\ }\href@noop {} {\emph {\bibinfo {title} {Many-Particle Physics}}}\
  (\bibinfo  {publisher} {Plenum},\ \bibinfo {address} {New York},\ \bibinfo
  {year} {1990})\BibitemShut {NoStop}%
\bibitem [{\citenamefont {Skinner}\ and\ \citenamefont
  {Shklovskii}(2013)}]{skinner_theory_2013}%
  \BibitemOpen
  \bibfield  {author} {\bibinfo {author} {\bibfnamefont {B.}~\bibnamefont
  {Skinner}}\ and\ \bibinfo {author} {\bibfnamefont {B.~I.}\ \bibnamefont
  {Shklovskii}},\ }\href {\doibase 10.1103/PhysRevB.87.075454} {\bibfield
  {journal} {\bibinfo  {journal} {Physical Review B}\ }\textbf {\bibinfo
  {volume} {87}},\ \bibinfo {pages} {075454} (\bibinfo {year}
  {2013})}\BibitemShut {NoStop}%
\bibitem [{\citenamefont {Kolomeisky}\ \emph {et~al.}(2013)\citenamefont
  {Kolomeisky}, \citenamefont {Straley},\ and\ \citenamefont
  {Zaidi}}]{kolomeisky_fermion_2013}%
  \BibitemOpen
  \bibfield  {author} {\bibinfo {author} {\bibfnamefont {E.~B.}\ \bibnamefont
  {Kolomeisky}}, \bibinfo {author} {\bibfnamefont {J.~P.}\ \bibnamefont
  {Straley}}, \ and\ \bibinfo {author} {\bibfnamefont {H.}~\bibnamefont
  {Zaidi}},\ }\href {\doibase 10.1103/PhysRevB.88.165428} {\bibfield  {journal}
  {\bibinfo  {journal} {Physical Review B}\ }\textbf {\bibinfo {volume} {88}},\
  \bibinfo {pages} {165428} (\bibinfo {year} {2013})}\BibitemShut {NoStop}%
\bibitem [{\citenamefont {Burkov}\ \emph {et~al.}(2011)\citenamefont {Burkov},
  \citenamefont {Hook},\ and\ \citenamefont
  {Balents}}]{burkov_topological_2011}%
  \BibitemOpen
  \bibfield  {author} {\bibinfo {author} {\bibfnamefont {A.~A.}\ \bibnamefont
  {Burkov}}, \bibinfo {author} {\bibfnamefont {M.~D.}\ \bibnamefont {Hook}}, \
  and\ \bibinfo {author} {\bibfnamefont {L.}~\bibnamefont {Balents}},\ }\href
  {\doibase 10.1103/PhysRevB.84.235126} {\bibfield  {journal} {\bibinfo
  {journal} {Physical Review B}\ }\textbf {\bibinfo {volume} {84}},\ \bibinfo
  {pages} {235126} (\bibinfo {year} {2011})}\BibitemShut {NoStop}%
\bibitem [{\citenamefont {{Abrikosov, Alexei
  A}}(1988)}]{abrikosov_alexei_a_fundamentals_1988}%
  \BibitemOpen
  \bibfield  {author} {\bibinfo {author} {\bibnamefont {{Abrikosov, Alexei
  A}}},\ }\href@noop {} {\emph {\bibinfo {title} {Fundamentals of the Theory of
  Metals}}}\ (\bibinfo  {publisher} {Elsevier},\ \bibinfo {address} {New
  York},\ \bibinfo {year} {1988})\BibitemShut {NoStop}%
\end{thebibliography}%

\end{document}